\documentclass[journal,12pt,onecolumn,draftclsnofoot,]{IEEEtran}

\usepackage[cmex10]{amsmath}
\usepackage[utf8]{inputenc}
\usepackage{multirow}
\usepackage{soul}
\usepackage{algorithm}
\usepackage{algpseudocode}
\usepackage{graphicx}
\usepackage{dsfont}
\usepackage{xcolor}
\usepackage{cite}

\usepackage{tabularray}

\usepackage[utf8]{inputenc}
\usepackage{mathtools}
\usepackage[mathscr]{euscript}
\usepackage{amsmath}
\usepackage{amssymb}
\usepackage{amsfonts}
\usepackage{amssymb}
\usepackage{amsmath}
\usepackage{verbatim}
\usepackage[T1]{fontenc}
\usepackage[utf8]{inputenc}
\usepackage{latexsym}
\usepackage{enumerate}
\usepackage{indentfirst}
\usepackage{calligra}
\usepackage{ulem}
\usepackage{graphicx}
\usepackage{array}
\usepackage{float}
\usepackage{bbm}
\usepackage{geometry}

\usepackage{mleftright}
\usepackage{colortbl}
\makeatletter
\let\old@ps@headings\ps@headings
\let\old@ps@IEEEtitlepagestyle\ps@IEEEtitlepagestyle
\def\psccfooter#1{%
 \def\ps@headings{%
 \old@ps@headings%
 \def\@oddfoot{\strut\hfill#1\hfill\strut}%
 \def\@evenfoot{\strut\hfill#1\hfill\strut}%
 }%
 \def\ps@IEEEtitlepagestyle{%
 \old@ps@IEEEtitlepagestyle%
 \def\@oddfoot{\strut\hfill#1\hfill\strut}%
 \def\@evenfoot{\strut\hfill#1\hfill\strut}%
 }%
 \ps@headings%
}
\makeatother

\usepackage{soul}
\usepackage{multirow}
\usepackage{soul,color}
\usepackage{graphicx}
\usepackage{dsfont}
\usepackage{subcaption}

\usepackage[utf8]{inputenc}
\usepackage{mathtools}
\usepackage[mathscr]{euscript}
\usepackage{amsmath}
\usepackage{amssymb}
\usepackage{amsfonts}
\usepackage{verbatim}
\usepackage[T1]{fontenc}
\usepackage[utf8]{inputenc}

\usepackage{latexsym}
\usepackage{enumerate}
\usepackage{indentfirst}
\usepackage{calligra}
\usepackage{ulem}
\usepackage{multirow}

\usepackage{array}
\usepackage{float}
\usepackage{bbm}

\usepackage{eurosym}

\usepackage{hyperref}

\usepackage{caption}
\usepackage{tikz}
\usepackage{pgfplots}

\pgfplotsset{compat=1.8}
\usepgfplotslibrary{statistics}
\pgfmathdeclarefunction{fpumod}{2}{%
 \pgfmathfloatdivide{#1}{#2}%
 \pgfmathfloatint{\pgfmathresult}%
 \pgfmathfloatmultiply{\pgfmathresult}{#2}%
 \pgfmathfloatsubtract{#1}{\pgfmathresult}%
 \pgfmathfloatifapproxequalrel{\pgfmathresult}{#2}{\def\pgfmathresult{5}}{}%
 }

\usepackage{tabularx}
\usepackage{flushend}
\usepackage{textcomp}
\usepackage{xcolor}
\usepackage{import}

\usepackage{csvsimple}
\usepackage{tabularx}
\usepackage{adjustbox}

\usetikzlibrary{trees,decorations,shadows}
\tikzset{level 1/.style={sibling angle=45,level distance=4mm}}
\usetikzlibrary{arrows.meta}
\usepackage{forest}
\usetikzlibrary{external}
\let\oldtikzexternalgetnextfilename\tikzexternalgetnextfilename \renewcommand{\tikzexternalgetnextfilename}[1]{\oldtikzexternalgetnextfilename{#1}\expandafter\tikzsetnextfilename\expandafter{#1}}

\usepackage{pgfplotstable}
\usepackage[outline]{contour}
\contourlength{1.2pt}

\pgfplotsset{compat=1.13} 

\usetikzlibrary{spy}
\usetikzlibrary{calc}
\usetikzlibrary{fadings}
\usetikzlibrary{patterns}
\usetikzlibrary{shadows}
\usetikzlibrary{mindmap}
\usetikzlibrary{backgrounds}
\usetikzlibrary{shapes.symbols}
\usetikzlibrary{shapes.multipart}
\usetikzlibrary{shapes.geometric}
\usetikzlibrary{automata,positioning}
\usetikzlibrary{decorations.fractals} 
\usetikzlibrary{decorations.markings}
\usetikzlibrary{decorations.pathreplacing}
\usetikzlibrary{decorations.pathmorphing}
\usepackage{svg}
 
\usepackage{bm}

\usepackage{nomencl}
\makenomenclature
\tikzset{edge from parent/.style={segment angle=10,draw}}

\tikzset{
 my rounded corners/.append style={rounded corners=2pt},
}

\def\BibTeX{{\rm B\kern-.05em{\sc i\kern-.025em b}\kern-.08em
 T\kern-.1667em\lower.7ex\hbox{E}\kern-.125emX}}

\renewcommand{\nomgroup}[1]{%
 \ifthenelse{\equal{#1}{O}}{\item[\textit{Operators}]}{%
 \ifthenelse{\equal{#1}{I}}{\item[\textit{Indices}]}{%
 \ifthenelse{\equal{#1}{A}}{\item[\textit{Acronyms}]}{%
 `\ifthenelse{\equal{#1}{V}}{\item[\textit{Variables and parameters}]}{}}}}}
\usepackage{scalerel}
\usetikzlibrary{svg.path}
\definecolor{orcidlogocol}{HTML}{A6CE39}
\tikzset{
 orcidlogo/.pic={
 \fill[orcidlogocol] svg{M256,128c0,70.7-57.3,128-128,128C57.3,256,0,198.7,0,128C0,57.3,57.3,0,128,0C198.7,0,256,57.3,256,128z};
 \fill[white] svg{M86.3,186.2H70.9V79.1h15.4v48.4V186.2z}
 svg{M108.9,79.1h41.6c39.6,0,57,28.3,57,53.6c0,27.5-21.5,53.6-56.8,53.6h-41.8V79.1z M124.3,172.4h24.5c34.9,0,42.9-26.5,42.9-39.7c0-21.5-13.7-39.7-43.7-39.7h-23.7V172.4z}
 svg{M88.7,56.8c0,5.5-4.5,10.1-10.1,10.1c-5.6,0-10.1-4.6-10.1-10.1c0-5.6,4.5-10.1,10.1-10.1C84.2,46.7,88.7,51.3,88.7,56.8z};
 }
}

\newcommand\orcidicon[1]{\href{https://orcid.org/#1}{\mbox{\scalerel*{ \begin{tikzpicture}[yscale=-1,transform shape]
 \pic{orcidlogo};
 \end{tikzpicture}
 }{|}}}}

\usepackage{bm}
\usepackage{mdframed}
\usepackage{lipsum}

\newmdenv[leftline=false,rightline=false,linewidth=1pt]{topbot}

\begin{document}
% \IEEEpubid{\begin{minipage}{\textwidth}\ \\[12pt]
%         \footnotesize{\\979-8-3503-9042-1/24/\$31.00 \copyright 2024 IEEE}
% \end{minipage}}
%
% paper title
\title{\huge{Linear energy storage and flexibility model with ramp rate, ramping, deadline and capacity constraints}}

\author{\IEEEauthorblockN{Md~Umar~Hashmi*\orcidicon{0000-0002-0193-6703},
~Dirk~Van~Hertem~\orcidicon{0000-0001-5461-8891}}
 
 \IEEEauthorblockA{\textit{KU Leuven \& EnergyVille},
Genk, Belgium}

 \IEEEauthorblockA{(mdumar.hashmi, dirk.vanhertem)@kuleuven.be}

 \and
 \IEEEauthorblockN{
Aleen van der Meer\orcidicon{0000-0003-3744-3238}, and~Andrew Keane\orcidicon{0000-0001-8527-514x}}
 
 \IEEEauthorblockA{\textit{University College Dublin,},
Dublin, Ireland.}

 \IEEEauthorblockA{aleen.vandermeer@ucdconnect.ie, andrew.keane@ucd.ie}

 }

% \author{Md~Umar~Hashmi*\orcidicon{0000-0002-0193-6703},
% Simon~Nagels\orcidicon{0009-0008-6487-6526},
% and~Dirk~Van~Hertem~\orcidicon{0000-0001-5461-8891}
% \thanks{Corresponding author email: mdumar.hashmi@kuleuven.be}
% \thanks{Md Umar Hashmi, Simon Nagels and Dirk Van Hertem are with KU Leuven, division Electa \& EnergyVille, Genk, Belgium}
% \thanks{This work is supported by 
% the \href{https://euniversal.eu/}{H2020 EUniversal project}, grant agreement ID: 864334 
% % (\url{https://euniversal.eu/}) 
% and 
% the Flemish Government and Flanders Innovation \& Entrepreneurship (VLAIO) through the IMPROcap project (HBC.2022.0733).
% }} 

\maketitle

\begin{abstract}
The power networks are evolving with increased active components such as energy storage and flexibility derived from loads such as electric vehicles, heat pumps, industrial processes, etc.
% water heaters, etc. 
% We will need better models representing these assets to harness the full potential of energy storage and flexibility. 
% There is a need for better models to represent these assets.
% Without such models, the true capability of such resources might be over or under-estimated. 
Better models are needed to accurately represent these assets; otherwise, their true capabilities might be over or under-estimated.
% In this work, we propose a new energy storage and flexibility energy arbitrage model that not only considers the ramp (power) and capacity (energy) limits but also accurately models the ramp rate constraint. 
In this work, we propose a new energy storage and flexibility arbitrage model that accounts for both ramp (power) and capacity (energy) limits, while accurately modelling the ramp rate constraint.
The proposed models are linear in structure and efficiently solved using off-the-shelf solvers as a linear programming problem. We also provide an online repository for wider application and benchmarking. Finally, numerical case studies are performed to quantify the sensitivity of ramp rate constraint on the operational goal of profit maximization for energy storage and flexibility. The results are encouraging for assets with a slow ramp rate limit. We observe that for resources with a ramp rate limit of 10\% of the maximum ramp limit, the marginal value of performing energy arbitrage using such resources exceeds 65\% and up to 90\% of the maximum profit compared to the case with no ramp rate limitations.
\end{abstract}

\begin{IEEEkeywords}
Energy storage, flexibility, ramp rate, linear programming, epigraph, energy arbitrage.
\end{IEEEkeywords}
 % \vspace{-10pt}

 \pagebreak

\tableofcontents

 \pagebreak

\section{Introduction}

The power networks' penetration of distributed energy resources (DER) is increasing. These DERs are weather-dependent power sources with a high simultaneity factor of power generation, making it challenging to balance the supply and demand of electricity. The infamous "duck curve" from California 
\cite{denholm2015overgeneration} necessitates quantifying flexibility adequacy not only in terms of ramping and energy capacity but also in ramp rates. In \cite{headley2020energy}, the ramp rate needed in California is considered to be 88.8 MW/min based on the 2017 DER level.

% \subsection{Importance of ramp rate modelling}
The importance of ramp rate modelling is highlighted in many prior works: 
\cite{cui2018estimating, coffman2020characterizing, holttinen2013flexibility, lannoye2012evaluation, correa2016dynamic, cornelius2014assessing, bahramara2021robust, zhao2024assessing, sanandaji2015ramping, adams2010flexibility, taibi2018power, yamujala2022enhancing, zhang2015impact}. 
Flexibility needs quantification is performed in \cite{zhao2024assessing, adams2010flexibility} considering the ramp rate limitations. 
\cite{correa2016dynamic} assesses the impact of ramp rate changes on the outcome of unit commitment.
Similar to unit commitment constraints, \cite{holttinen2013flexibility} also considers the minimum up/down time and start-up time along with the ramp rate constraint. For flexibility and storage models in this work, the maximum uptime
% and downtime 
is inherently constrained by the capacity or quality-of-service, as in \cite{coffman2020characterizing}. 
However, the downtime and start-up time 
constraints are ignored. 
% Although, for generators, this cannot be ignored. 
Application or device-specific flexibility models for smart homes with DER, HVAC loads, and electric vehicles considering ramp rate constraints are detailed in \cite{cui2018estimating, sanandaji2015ramping, zhang2015impact} respectively.

% \textcolor{red}{
% Some real-world motivations for performing ramp rate modelling for storage devices. These storage devices are of many types: batteries including lead acid and Li-Ion, flywheel, pump hydro, potential energy buffers etc.
% and flexibility (derived from heat pumps, water heaters, HVAC etc.).
% For storage check out energyvault.com
% }

\subsection{Motivation for ramp rate constraint modelling}
Modelling ramp rate constraints is crucial for flexibility and storage modelling in power system analysis for several reasons. These constraints ensure the reliable and efficient operation of the power grid by addressing the dynamic behavior of various energy resources. The key reasons motivating the modelling of ramp rate constraints are:
\begin{itemize}
\item \textit{Integration of Renewable Energy}: Renewable energy sources like solar and wind are intermittent and unpredictable. Ramp rate constraints help in managing the integration of these sources by ensuring that the grid can smoothly handle their variability. As renewable energy generation can change rapidly (e.g., clouds covering a solar farm), ramp rate constraints ensure that other sources providing resource adequacy can ramp up or down to maintain a balance \cite{denholm2011grid}.
\item \textit{Grid Stability}: The ramp rate constraints help maintain grid frequency within acceptable limits by ensuring that power injections or withdrawals from flexible resources and storage devices can respond quickly enough to sudden changes in demand or supply. Rapid changes in power output can cause voltage fluctuations, potentially destabilizing the grid \cite{kirby2005frequency}. Ramp rate constraints help manage these fluctuations.
\item \textit{Economic Dispatch and Unit Commitment}: Ramp rate constraints are essential for the economic dispatch and unit commitment processes, ensuring that generation and storage units are scheduled to minimize costs while meeting demand and operational limits. Non-compliance with ramp rate constraints can lead to penalties or higher operational costs due to the need for emergency measures.
\end{itemize}

Ramp rate constraints for sources of flexibility are bounded as:
\begin{itemize}
\item \textit{Pumped hydro plants} have significant ramping capabilities but must adhere to ramp rate limits to avoid mechanical stress and coordinate effectively with other grid resources. %For example, if the grid requires a quick ramp-up in power supply, PHS can respond, but within the limits of its ramp rate.
\item \textit{Batteries} can respond quickly to power demands, making them ideal for frequency regulation. However, their ramp rate is constrained by the battery chemistry and its thermal limits. Modelling these constraints ensures that the battery operation is within safe limits and enhances grid reliability.
\item \textit{Flexible devices} such as HVAC: the flexibility provided by loads is constrained by rated power, exogenous parameters such as temperature and humidity and the quality of service to be provided. These constraints would limit the ramp rate provided by such flexibility sources. If not considered, it would lead to over-estimation of the amount of flexibility.
\end{itemize}

\begin{table}
\centering
\caption{Parameter for storage and flexible sources}
\scriptsize
\begin{tblr}{
  hlines,
  vlines,
}
\textbf{Resource}             & \textbf{Ref} & \textbf{Power} & \textbf{Energy} & \textbf{Ramp Rate} \\
G-vault (DC)        &   \cite{energyvault}           & 6.25MW         & 4-2 hours & 0-100\% in 120sec  \\
PhotoVoltaic (PV)                            &       \cite{marcos2011power}         & -              & -               & 90\% in 20s        \\
Molten salt storage    &     \cite{moltensalt}           & 100 MW         & -               & 10\% per min       \\
Pumped hydro  &   \cite{pumpedhydro}             & 200-250 MW     & 10 GWh          & 0-100\% in 2 min   \\
Flywheel  &     \cite{Stornetic}           & 130 kW     & 4.1 kWh         & 20 ms   
% \\
% Inverter &      \cite{inverterArena}       & -              & -               & 16.67\% /min        
\end{tblr}
\label{tab:paraLit}
\end{table}

Tab. \ref{tab:paraLit} details some commercial storage and flexibility parameters. This list shows the variation of the parameters. On the one hand, we have fast ramp rate devices such as flywheel, batteries that can sustain the power output for a limited period. For example, the Stornetic's EnWheel 130 has one directional capacity, lasting for just 50 seconds \cite{Stornetic}.
While molten storage and pumped hydro could be examples of relatively slow ramp-rate storage devices.

\pagebreak

\subsection{Contributions of the paper}

The contributions of the paper are as follows:\\
$\bullet$ \textit{Energy storage and flexibility model with ramp rate constraint}: we extend the flexibility model proposed in \cite{hashmi2022can} and the energy storage model proposed in \cite{hashmi2019optimal} with the ramp rate constraint. The flexibility model also includes a desired time window of operation and a deadline constraint, ensuring the energy consumed is temporally optimized within this time window. The proposed storage model also considers ramp rate limitations along with ramp and capacity constraints. These models are linear and can be efficiently solved using off-the-shelf solvers. 
    The run-time for 1000 Monte Carlo (MC) simulation days with 96 time steps for each MC scenario takes less than 30 seconds. Due to its computational efficiency, the proposed models are suitable for (near) real-time operation. \\
$\bullet$ \textit{Numerical evaluations}: two case studies are presented showcasing the efficacy of the models. In the first case study, the battery model is evaluated. It is observed that the marginal value of energy storage arbitrage profit is quite high for smaller ramp rate values. This is an encouraging finding for low (compared to ramp limits) ramp-rate storage devices. 
    In the second case study, it is observed that flexible load can be temporally shifted. The inclusion of the ramp-rate constraint substantially reduced the switching of the flexible device for a small marginal increase in the reduction of energy consumption cost, similar to \cite{hashmi2018limiting}.\\
$\bullet$ \textit{Online repository}: 
For wider applicability and quicker benchmarking storage and flexibility models are provided in an online repository in three different software-compatible codes in Matlab, Python and GAMS. The repository is publicly available here: \url{https://github.com/umar-hashmi/FlexibilityAndStorageModel}

This paper is organized as follows:
Section \ref{section2} details the energy storage and flexibility models.
Section \ref{section3} describes the linear programming formulation for utilizing storage and flexibility models for energy arbitrage.
Section \ref{section4} describes the numerical case studies for assessing the proposed framework.
Section \ref{section5} concludes the paper.

\pagebreak

\section{Energy storage and flexibility models}
\label{section2}
The energy storage and flexibility models often use three parameters for defining operational constraints, i.e., (a) ramp rate, (b) power, and (c) energy \cite{ulbig2015analyzing, bucher2015quantification}.
The units used for ramp rate, power and energy are watt per second, watt, and joule, respectively.
% The metrics for describing storage and flexible resources use units such as watt per second, watt, and joule.
Flexible resources can be categorized into ramp-up and ramp-down flexibility. Ramp-down flexibility refers to the reduction of nodal load akin to load curtailment, while ramp-up flexibility increases nodal load similar to generation curtailment. Ramp-down flexibility is achieved by deactivating consumer loads like HVAC systems, water heaters, and pool pumps \cite{chen2018distributed}. Conversely, ramp-up flexibility can be attained through solar generation curtailment and/or activation of electrical loads.
These flexible resources are subject to limitations in ramp rate, power, and energy that dictate their utilization as flexibility. For instance, the temperature of a water heater may be constrained within certain bounds. If employed as ramp-up flexibility to boost nodal load, it can only operate until the water temperature reaches the upper bound (or lower bound for ramping down). Therefore, to define flexible resources, we incorporate constraints on power, energy, and ramp rate to uphold technical device specifications and user comfort requirements \cite{hashmi2023robust}.

On the other hand, prosumer energy storage if not fully charged or discharged can provide both ramp-up and ramp-down flexibility \cite{buvsic2017distributed}.
Next, we present the improved storage and flexibility model that includes the ramp rate constraint for both storage and flexibility and the deadline constraints for flexible resources \cite{subramanian2012real}. The deadline constraint is crucial to ensure the quality of service \cite{oikonomou2020coordinated}.

\pagebreak

\subsection{Energy storage model}
% In the storage model, we account for ramp rate (unit of power per unit time), ramping constraint (unit of power), and capacity constraint (unit of energy), along with charging and discharging efficiencies, represented by \( \eta_{\text{ch}} \) and \( \eta_{\text{dis}}  \in (0,1] \) respectively. The energy optimization involves considering the change in energy levels of the storage at time \( i \), denoted as \( x_i \).
% Change in storage energy level at $i$ is defined as $x_i = h \delta_i$, where $\delta_i \in [\delta_{\min}, \delta_{max}]$ $\forall i$ denotes ramp limit of the storage. 
% $h$ denotes the sampling period.
% $\delta_i> 0 $ when the storage is charging and vice versa. Note, $\delta_i$ is in units of power (MW) and $x_i$ is in units of energy (MWh). 
% For time $i$, the storage charge level is denoted as

In the storage model, we consider the ramp rate (power per unit time), ramping constraint (power), and capacity constraint (energy). Additionally, charging and discharging efficiencies, represented by \( \eta_{\text{ch}} \) and \( \eta_{\text{dis}} \) respectively, are taken into account, where \( \eta_{\text{ch}}, \eta_{\text{dis}} \in (0,1] \). The energy optimization process involves analyzing the change in energy levels of the storage at time \( i \), denoted as \( x_i \).

The change in storage energy level at time \( i \) is given by \( x_i = h \delta_i \), where \( \delta_i \in [\delta_{\min}, \delta_{\max}] \) for all \( i \) represents the ramp limit of the storage. Here, \( h \) denotes the sampling period. When \( \delta_i > 0 \), the storage is charging, and when \( \delta_i < 0 \), the storage is discharging. Note that \( \delta_i \) is measured in units of power (MW), and \( x_i \) is measured in units of energy (MWh). 

At time \( i \), the storage charge level is represented as follows:
\begin{equation}
b_i = b_{i-1} + x_i, \quad b_i\in [b_{\min},b_{\max}], \forall i,
\label{eq:Batcap}
\end{equation}
where $b_{\min}, b_{\max}$ are the minimum and maximum storage capacity within which the storage should be operating. The power consumed by the storage at time $i$ is denoted as
\begin{equation}
f(x_i)\text{= } \frac{[x_i]^+}{h \eta_{\text{ch}}} - \frac{\eta_{\text{dis}}[x_i]^-}{h}\text{= }\frac{\max(0,x_i)}{h\eta_{\text{ch}}} - \frac{\eta_{\text{dis}}\max(0,-x_i)}{h}, 
\label{finverse}
\end{equation}
where $x_i$ must lie in the range from $X_{\min}=\delta_{\min}h$ to $X_{\max}={\delta_{\max}h}$.
% The total power consumed between time step $i$ and $i+1$ is given as $L_i = z_i+y_i +f(x_i)$, where $z_i$ is the uncontrolled net load, $y_i$ is the flexible load and $f(x_i)$ is the battery consumption. 
The ramping constraint is given as
\begin{equation}
    x_i \in [X_{\min}, X_{\max}].
    \label{eq:BatRamp}
\end{equation}

Storage systems are commonly connected through a converter, which can be either an inverter or a rectifier. Let the efficiency of this converter be represented by \( \eta_{\text{conv}} \) where \( \eta_{\text{conv}} \in (0,1] \). The efficiency of charging and discharging the storage, modified due to this converter, is denoted as
\begin{equation}
    \eta_{\text{ch}}^* = \eta_{\text{ch}}\eta_{\text{conv}},~~~~
    \eta_{\text{dis}}^* = \eta_{\text{dis}}\eta_{\text{conv}}.
    \label{eq:eff}
\end{equation}

\eqref{eq:eff}
showcasing the modified storage efficiency
encapsulates the converter efficiency \cite{hashmi2023multi}.

The ramp rate constraint for the storage is denoted as
\begin{equation}
    x_i - x_{i-1} \in [\tau_{\min}, \tau_{\max}].
    \label{eq:BatRampRate}
\end{equation}
The ramp rate limits for storage are denoted as $\tau_{\min}, \tau_{\max}$ are minimum and maximum bounds. Further, the ramp rate limit is at best equal to the ramp limit given as
\begin{equation}
    \tau_{\min} \geq X_{\min}, ~~ \tau_{\max} \leq X_{\max}.
\end{equation}

\pagebreak

\subsection{Flexibility model}
The flexibility model is initially introduced in \cite{hashmi2022can}. 
In contrast to the storage model, which encompasses (a) ramp, (b) ramp rate, and (c) capacity constraints applicable at all times, the flexibility model includes (i) ramp, (ii) ramp rate, and (iii) a deadline constraint. While constraints (i) and (ii) apply at every time instance, constraint (iii) is relevant only at the end of the flexibility operation time window. Generally, constraint (iii) is an equality constraint, but we implement it as an epsilon inequality constraint, see \eqref{eq:loadflexibility}.
The flexible portion of the load can be managed within certain limits while ensuring that the total energy consumed does not decrease. This deadline constraint is expressed as follows:
% The flexibility model is first introduced in \cite{hashmi2022can}.
% Unlike the storage model that consists of (a) ramp, (b) ramp rate and (c) capacity constraint applicable for all times. The flexibility model consists of (i) ramp, (ii) ramp rate and (iii) deadline constraint. Although, constraints (i) and (ii) are applicable for each time, while (iii) is only applicable for the end of the flexibility time window of operation. The constraint (iii) is typically an equality constraint that we used as an epsilon inequality constraint.
% The flexible component of the load can be controlled within a range while ensuring the cumulative energy consumed is not decreased. This deadline constraint is given by
\begin{equation}
 K - \epsilon \leq h \sum_{i=t_a}^{t_d} y_i \leq K +\epsilon, \forall i,
\label{eq:loadflexibility}
\end{equation}
\begin{equation}
y_i \in [y_{\min}^i,y_{\max}^i], ~ \forall i.
\label{eq:loadflexibility2}
\end{equation}
In this context, \( K \) denotes the cumulative energy consumption goal for flexible loads, which must be met while maintaining service quality. The operational flexibility window is determined by \( t_a \), the start (arrival) time, and \( t_d \), the end (departure) time. Flexible loads operate within the bounds set by the upper and lower limits, \( y_{\max}^i \) and \( y_{\min}^i \), respectively. Because these loads (i.e. consuming electrical energy) provide flexibility, it follows that \( y_{\min}^i \geq 0 \). Equation \eqref{eq:loadflexibility} guarantees that the benefits from arbitrage come from effective energy management rather than a mere reduction in total energy usage. The term \( \epsilon \) is a small parameter that ensures the total energy used by flexible loads is nearly equal to the target level of \( K \), allowing for a small margin of error. Without this margin, Equation \eqref{eq:loadflexibility} turns into an equality constraint.

The inequality constraint is changed into an equality constraint by using slack variable \cite{madsen2004optimization}, while the equality constraint is transformed to an inequality constraint using epsilon-neighborhood \cite{si2014equality, blogEpsilon}.

The ramp rate constraint for flexibility is denoted as
\begin{equation}
    y_i - y_{i-1} \in [\xi_{\min}, \xi_{\max}].
    \label{eq:flexRampRate}
\end{equation}
The ramp rate limits for flexibility are denoted as $\xi_{\min}, \xi_{\max}$ are minimum and maximum bounds. Further, the ramp rate limit is at best equal to the ramp limit given as
$
    \xi_{\min} \geq y_{\min}, ~~ \xi_{\max} \leq y_{\max}.
$

\pagebreak

\section{Optimization model for flexibility and storage}
\label{section3}

The goal of storage and flexibility is to perform energy arbitrage based on the time-varying price of electricity.
Similar to \cite{hashmi2017optimal, hashmi2019optimal}, the selection of $x_i$ as the decision variable ensures that the arbitrage problem is convex, provided {the ratio of selling and buying price of electricity denoted as $\kappa$ satisfies} $\kappa \leq 1$.  

The proposed storage and flexibility energy arbitrage models are detailed in Sec. \ref{sub1model1} and \ref{sub2model2}, respectively. These models include an efficient model for the ramp rate constraint that was not considered in prior works \cite{hashmi2019optimal, hashmi2019optimization}.

\pagebreak

\subsection{Model 1: Storage model with ramp rate constraint}
\label{sub1model1}
The storage model for arbitrage is formulated as a linear programming-based problem using the epigraph equivalent formulation for piecewise linear convex cost function. %, we formulate the optimal arbitrage problem using LP. 
% The associated linear programming matrices can be found in \cite{hashmi2019optimization, hashmi2019optimal}.
The proposed optimization problem, $P_{\text{Bat}}$, considers the ramp rate constraint, unlike prior works \cite{hashmi2019optimization, hashmi2019optimal}, on which this paper builds upon. The proposed storage optimization model is given as

\begin{topbot}
    \text{Proposed battery model with ramp rate constraint ($P_{\text{Bat}}$)} 
\end{topbot}
{\allowdisplaybreaks
\begin{IEEEeqnarray}{ l C  }
    \label{eq:epi_formulation}
    \text{Objective function:} & \min \quad \{t_1 + t_2+...+t_N\} \nonumber \\
    % \min_X f^T X, &~& \\
    \text{subject to:} &\text{Segment 1:~} {p_b^i} x_i/{\eta_{ch}^*} - t_i \leq 0, \forall~ i,  \label{eqBatSeg1} \\ 
 &\text{Segment 2:~} {p_s^i}{\eta_{dis}^*} x_i - t_i \leq 0, \forall~ i, \label{eqBatSeg2} \\
    & \eqref{eq:Batcap}, \eqref{eq:BatRamp}, \eqref{eq:BatRampRate}. \nonumber \\
    % A.X \leq b, &~& \\
    % X \in [lb, ub], &~&  \\
    \hline \nonumber
\end{IEEEeqnarray}
}

\eqref{eqBatSeg1} and \eqref{eqBatSeg2} denote the equations for the two segments of the piecewise linear objective function for each time $i$.
A generic LP problem has an objective function denoted as $\min_X f^T X$, subject to inequality constraint denoted as $A.X \leq b$ and limit bounds on the decision variable given as $X \in [lb, ub]$.
The matrix format for the above problem is denoted as 
\begin{equation}
f\text{=}{\begin{bmatrix}
	0\\
	1\\
	\end{bmatrix}}_{2N\times 1},
X \text{=} {\begin{bmatrix}
	x_i\\
	t_i\\
	\end{bmatrix}}, 
lb\text{=}
{\begin{bmatrix}
	X_{\min}\\
	T_{\min}\\
	\end{bmatrix}}
\leq
{\begin{bmatrix}
	x_i \\
	t_i \\
	\end{bmatrix}} \leq
ub\text{=}
{\begin{bmatrix}
	X_{\max}\\
	T_{\max}\\
	\end{bmatrix}}.
\label{mateqsame}
\end{equation}
where $T_{\min}$ and $T_{\max}$ are bounds on $t_i$. Since these bounds are not known to us, we choose $T_{\min}$ to be negative with a large magnitude and $T_{\max}$ to be positive with a large magnitude.
%The dimension of $A_{c2}$ is 4Nx2N, $b_{c2}$ is 4Nx1, $X$ and $f$ is 2Nx1.

The LP formulation for \textit{A} and \textit{b} matrices are detailed below:

\begin{equation}
\footnotesize
{A}_{\text{bat}} \text{= }{ 
\renewcommand\arraystretch{1.3}
\mleft[ \begin{array}{c |c }
	\texttt{diag}_N(\frac{p_b}{\eta_{ch}^*}) & - \texttt{diag}_N \\
        \texttt{diag}_N({p_s}{\eta_{dis}^*}) & - \texttt{diag}_N \\
        \texttt{trigL}_N & \texttt{zeros}_N \\
        -\texttt{trigL}_N & \texttt{zeros}_N \\
        \texttt{DiffMt}_N & \texttt{zeros}_N \\
        -\texttt{DiffMt}_N & \texttt{zeros}_N \\
\end{array} \mright]}, 
\quad
{b}_{\text{bat}} \text{= }{\begin{bmatrix}
	 \\
	\texttt{zeros}_{(N,1)}\\
        \texttt{zeros}_{(N,1)}\\
        (b_{\max} - b_0)\texttt{ones}_{(N,1)}\\
        (b_0- b_{\min})\texttt{ones}_{(N,1)}\\
        X_{\max}\\
	\tau_{\max}.\texttt{ones}_{(N-1,1)}\\
        -X_{\min}\\
	-\tau_{\min}.\texttt{ones}_{(N-1,1)}\\
	\end{bmatrix}},\\
\end{equation} 

here $\texttt{diag}_N$ denotes a diagonal matrix of size N, $\texttt{trigL}_N$ denotes the lower triangular matrix of size N, $\texttt{ones}_{(x,y)}$ and $\texttt{zeros}_{(x,y)}$ denotes a matrix of all 1s and 0s with $x$ rows and $y$ columns respectively.

$\texttt{DiffMt}_N$ is defined as
\begin{equation}
\footnotesize
\texttt{DiffMt}_N = { 
\renewcommand\arraystretch{1.3}
\mleft[ \begin{array}{c c c  c c c}
	1 & 0 & 0 & .. & 0 & 0\\
        -1 & 1 & 0 & .. & 0 & 0\\
        0& -1 & 1 &  .. & 0 & 0\\
        : & : & : & .. & : & :\\
        0 & 0 & 0 & .. & -1 & 1\\
\end{array} \mright]}_{N\times N}, 
\end{equation}

Although the proposed formulation for energy storage arbitrage, $P_{\text{Bat}}$, does not assume inelastic load consumption, this model can be extended to consider an uncontrollable portion as proposed in \cite{hashmi2019optimal}.

\pagebreak

\subsection{Model 2: Flexibility model with ramp rate constraint}
\label{sub2model2}

In our earlier work \cite{hashmi2019optimal}, we {used} a linear programming (LP) based formulation to solve the optimal energy storage arbitrage problem. This LP formulation uses epigraph-based minimization \cite{boyd2004convex}. 
We extend the LP formulation in \cite{hashmi2019optimal} for including control of flexible load under time-varying electricity price. The formulation in \cite{hashmi2019optimal} uses the geometry of the cost function to form 4 line segments over which the epigraph-based cost function is minimized. Since flexible load $y_i \geq 0$, the operation of flexible loads does not lead to adding more line segments to the geometry of the cost function in \cite{hashmi2019optimal}. The LP formulation for solving the arbitrage problem is given as $P_{\text{Flex}}$.

\begin{topbot}
    \text{Proposed flexibility model with ramp rate constraint ($P_{\text{Flex}}$)}
\end{topbot}
{\allowdisplaybreaks
\begin{IEEEeqnarray}{ l C  }
    \label{eq:epi_formulation}
    \text{Objective function:} & \min \quad \{t_1 + t_2+...+t_N\} \nonumber \\
    % \min_X f^T X, &~& \\
    \text{subject to:} & \text{Segment 1:~}p_b^i y_i- t_i \leq 0, ~\forall~ i,  \label{eqFlexSeg1} \\
    & \text{Segment 2:~} p_s^i y_i - t_i \leq 0, ~\forall~ i, \label{eqFlexSeg2} \\
    & \eqref{eq:loadflexibility}, \eqref{eq:loadflexibility2}, \eqref{eq:flexRampRate}. \nonumber \\
    % A.X \leq b, &~& \\
    % X \in [lb, ub], &~&  \\
    \hline \nonumber
\end{IEEEeqnarray}
}

The matrix format for the optimization problem $P_{LP}$ is denoted as 
minimize ${f}^T X$, subject to ${A}X\leq b$, and $X \in [lb, ub]$.
% The dimension of $A$ is (6N+2)x3N, $b$ is (6N+2)x1, $X$ and $f$ are of size 3Nx1, N denotes number of samples in the horizon of optimization.

\begin{equation}
f\text{=}{\begin{bmatrix}
	0\\
	1\\
	\end{bmatrix}},
X \text{=} {\begin{bmatrix}
	y_i\\
	t_i\\
	\end{bmatrix}},
\label{mateqsame0}
\end{equation}
\vspace{-10pt}
\begin{equation}
lb\text{=}
{\begin{bmatrix}
	y_{\min}(i)\\
	T_{\min}\\
	\end{bmatrix}}
\leq
{\begin{bmatrix}
	y_i \\
	t_i \\
	\end{bmatrix}} \leq
ub\text{=}
{\begin{bmatrix}
	y_{\max}(i)\\
	T_{\max}\\
	\end{bmatrix}}.
\label{mateqsame}
\end{equation}
where $T_{\min}$ and $T_{\max}$ are bounds on $t_i$. Since these bounds are not known to us, we choose $T_{\min}$ to be negative with a large magnitude and $T_{\max}$ to be positive with a large magnitude.

The A and b matrices for flexibility are given as
\begin{equation}
\footnotesize
{A}_{\text{flex}} \text{= }{ 
\renewcommand\arraystretch{1.3}
\mleft[ \begin{array}{c |c }
	\texttt{diag}_N({p_b}) & -\texttt{diag}_N \\
        \texttt{diag}_N({p_s}) & -\texttt{diag}_N \\
        \texttt{row}_N(t_a,t_d) & \texttt{zeros}_N \\
        -\texttt{row}_N (t_a,t_d) & \texttt{zeros}_N \\
        \texttt{DiffMt}_N(t_a,t_d) & \texttt{zeros}_N \\
        -\texttt{DiffMt}_N(t_a,t_d) & \texttt{zeros}_N \\
\end{array} \mright]}, 
\quad
{b}_{\text{bat}} \text{= }{\begin{bmatrix}
	 \\
	\texttt{zeros}_{(N,1)}\\
        \texttt{zeros}_{(N,1)}\\
        K + \epsilon\\
        -K+\epsilon\\
        Y_{\max}\\
	\xi_{\max} .\texttt{ones}_{(N-1,1)}\\
        -Y_{\min}\\
	-\xi_{\min}.\texttt{ones}_{(N-1,1)}\\
	\end{bmatrix}}\\
\end{equation}

% %\pagebreak

\pagebreak

\section{Numerical results}
\label{section4}

\begin{figure}[!htbp]
    \centering
    \includegraphics[width=14cm]{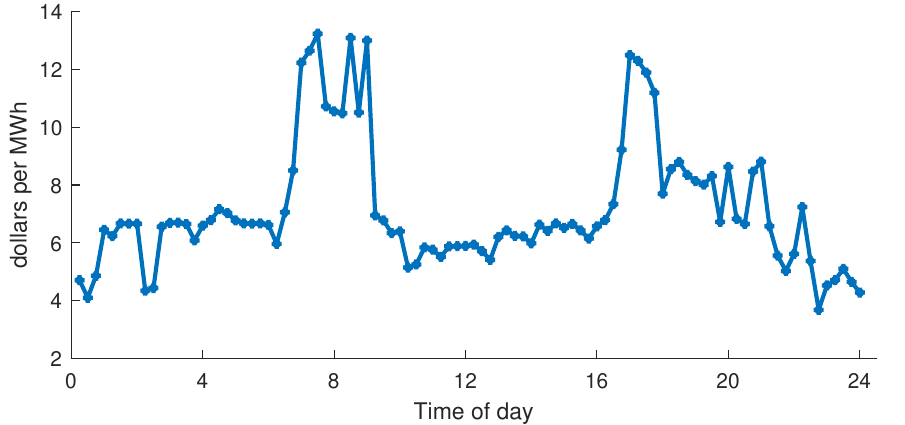}
    \vspace{-4pt}
    \caption{\small{Real-time electricity price signal.}}
    \label{fig:case0}
\end{figure}

Numerical results for performing energy arbitrage for storage and flexibility based on time-varying electricity prices. The price signal considered for this numerical evaluation is shown in Fig. \ref{fig:case0}. The price signal is based on real-time electricity prices from NYISO \cite{hashmi2017optimal}.

The numerical evaluation consists of two case studies. In these case studies, the proposed storage and flexibility models with ramp rate constraints are evaluated. In these evaluations, the impact on the storage model operation due to different levels of storage ramp rate is quantified.

\pagebreak

\subsection{Case study 1: storage optimization}
The energy storage considered for this case has the following parameters: $\eta_{\text{ch}}^*=\eta_{\text{dis}}^* = 0.95$, $b_{\max} = 1000$ watt-hour, $b_{min}=200$ watt-hour. 

\begin{figure}[!htbp]
    \centering
    \includegraphics[width=15.6cm]{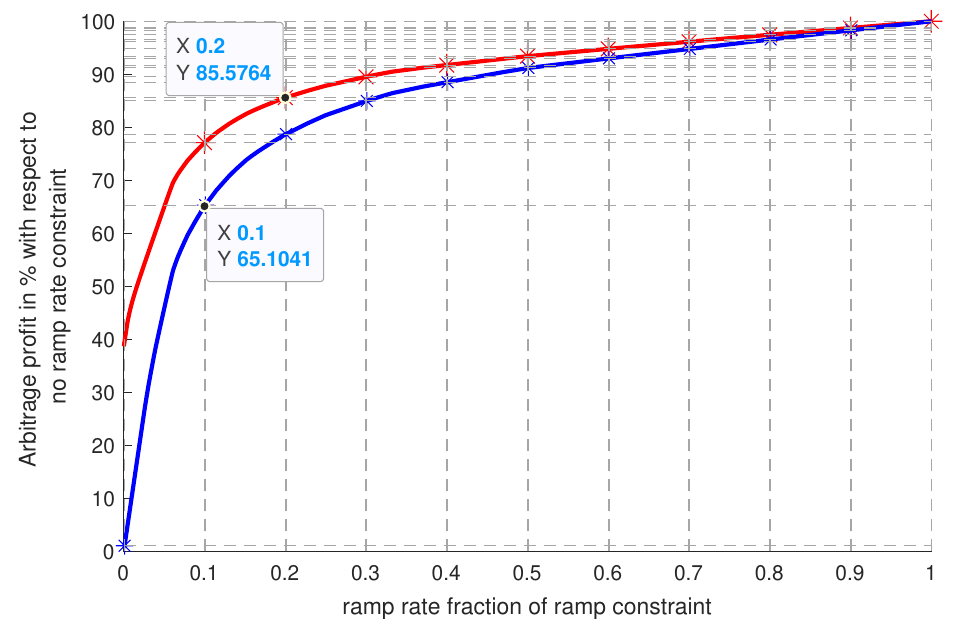}
    \vspace{-4pt}
    \caption{\small{Marginal arbitrage gain plot for $b_0=b_{\min}$ shown in blue and $b_0=b_{\max}$ shown in red.}}
    \label{fig:case1_res1}
\end{figure}

\textit{Sensitivity of the ramp rate:}
The impact of the initial battery charge is shown on the marginal arbitrage gain with different levels of ramp rate. The ramp rate constraint is shown as a fraction of the maximum ramping capability. For this fraction equal to 1 implies there is no ramp rate constraint. 
Fig. \ref{fig:case1_res1} shows that even if the ramp rate is 10\% of the maximum ramp capability, the storage can extract up to 65 and 78\% of the maximum arbitrage gains for $b_0=b_{\min}$ and $b_0=b_{\max}$, respectively.
This is a promising result for storage devices with slow ramp rate limits.

\begin{figure}[!htbp]
    \centering
    \includegraphics[width=15.5cm]{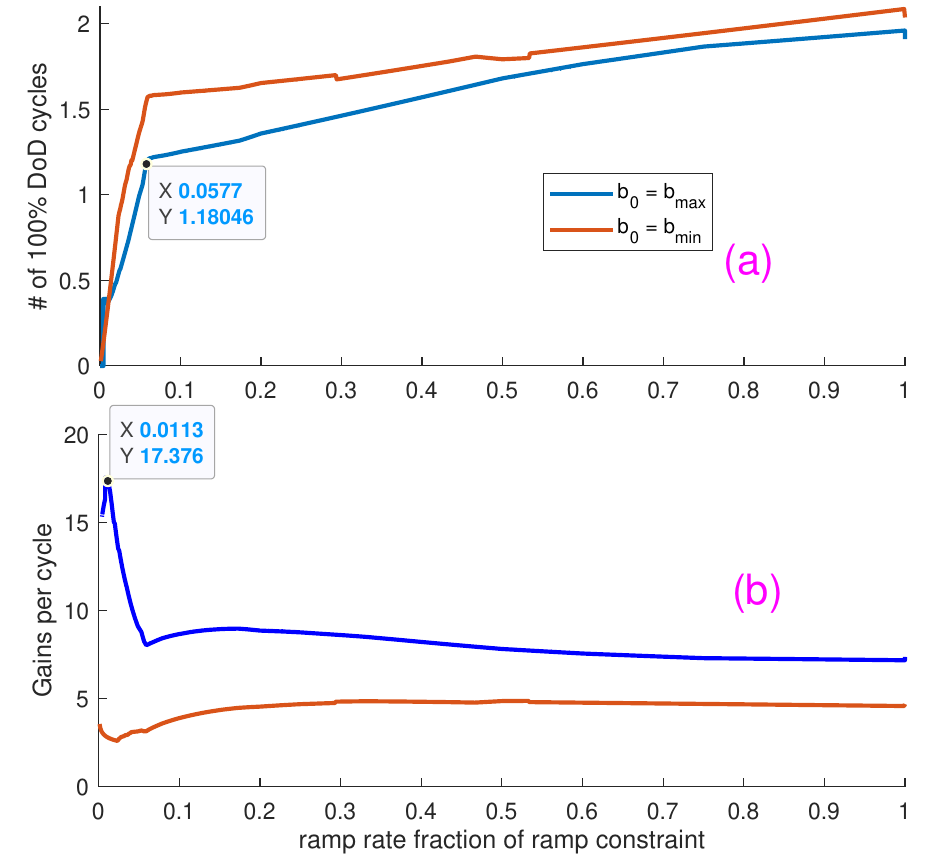}
    \vspace{-4pt}
    \caption{\small{Sensitivity of ramp rate constraint on (a) number of cycles of operation and (b) arbitrage gains per cycle for $b_0=b_{\min}$ shown in blue and $b_0=b_{\max}$ shown in red.}}
    \label{fig:case1_res2}
\end{figure}

The 100\% cycles of battery operation are counted based on \cite{hashmi2018long}. Fig. \ref{fig:case1_res2}(a) showcases the number of 100\% depth-of-discharge (DoD) cycles performed by the storage device. 
Fig. \ref{fig:case1_res2}(b) presents the arbitrage gains per 100\% DoD cycle of operation.

\begin{figure}[!htbp]
    \centering
    \includegraphics[width=16.5cm]{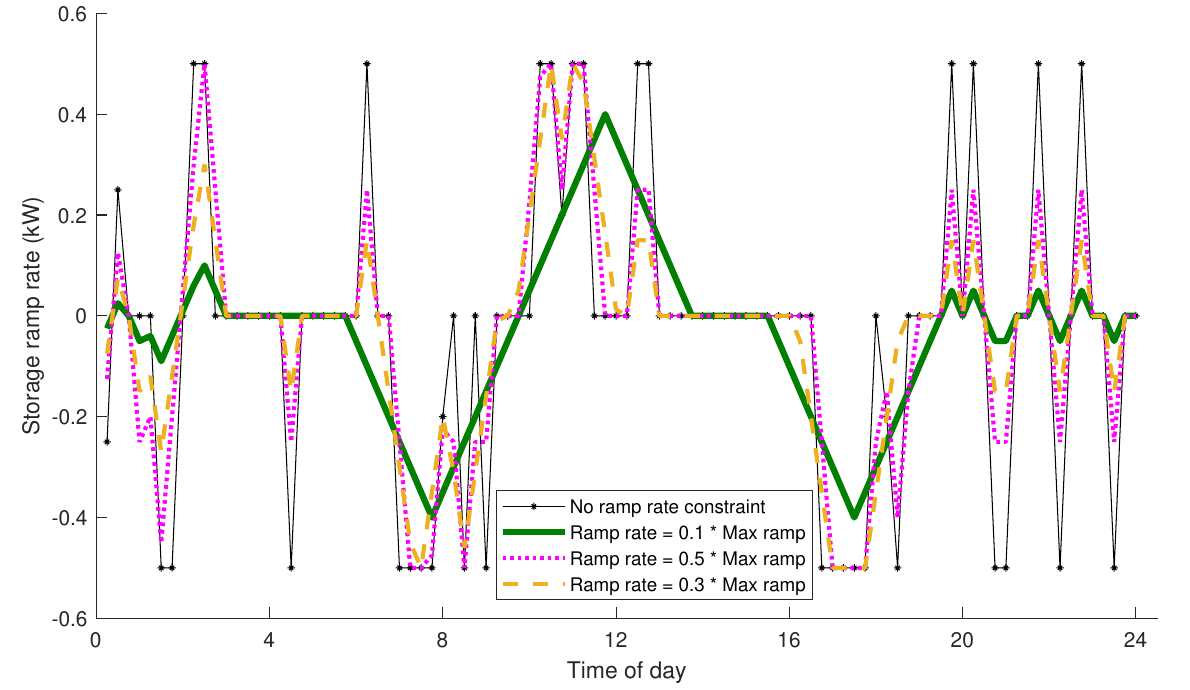}
    \vspace{-4pt}
    \caption{\small{Energy storage power output for max ramp rate set at 0.5 kW for $b_0=b_{\max}$.}}
    \label{fig:case1_res3}
\end{figure}

Fig. \ref{fig:case1_res3} shows the power output of the energy storage for different levels of ramp rate constraint with respect to the maximum ramp limit of 0.5 kW.

\begin{figure}[!htbp]
    \centering
    \includegraphics[width=15cm]{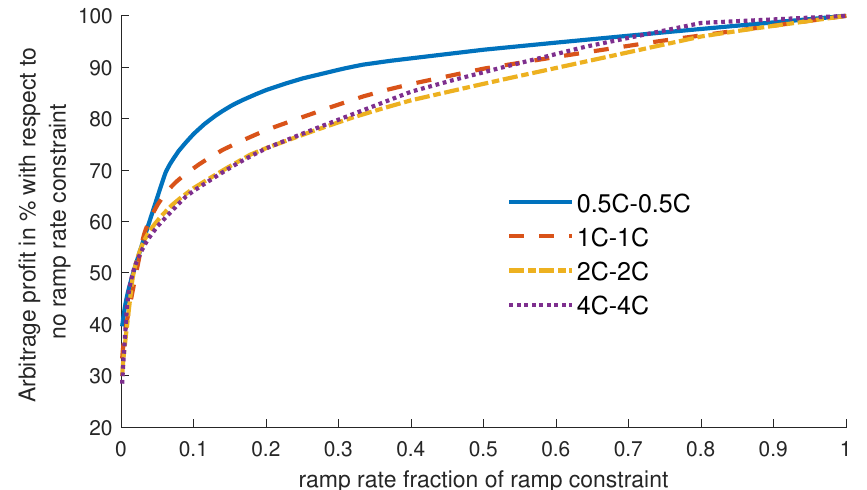}
    \vspace{-4pt}
    \caption{\small{Marginal arbitrage gains with different ramp levels of energy storage devices along with varying ramp rate limit.}}
    \label{fig:case1_res4}
\end{figure}

\textit{Sensitivity of ramp limit}: We denote the relationship between ramp rate and battery capacity using the xC-yC notation. In this notation, xC-yC signifies that the battery requires 1/x hours to charge fully and 1/y hours to discharge completely.
The ramp rate constraint has a similar impact on slow or fast ramping energy storage, refer to Fig. \ref{fig:case1_res4}. However, the impact on slower ramping storage is more pronounced for slower ramp rate limits.

\begin{figure}[!htbp]
    \centering
    \includegraphics[width=15cm]{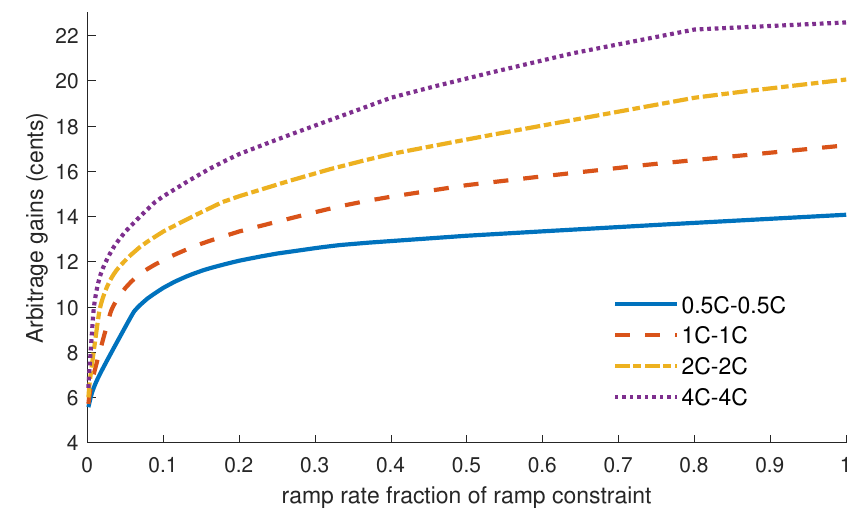}
    \vspace{-4pt}
    \caption{Aggregated arbitrage gains.}
    \label{fig:case1_res5}
\end{figure}

\textit{Computation time}: Simulations are performed on HP Intel(R) Core(TM) i7 CPU, 1.90GHz, 32 GB RAM personal computer on Matlab 2021a.
1000 Monte Carlo day simulations, with each day having 96 samples, take the storage model to solve within 30 seconds. Thus, simulating 1 day on average took <~0.03 seconds.

\pagebreak

\subsection{Case study 2: flexibility optimization}
\begin{figure}[!htbp]
    \centering
    \includegraphics[width=13cm]{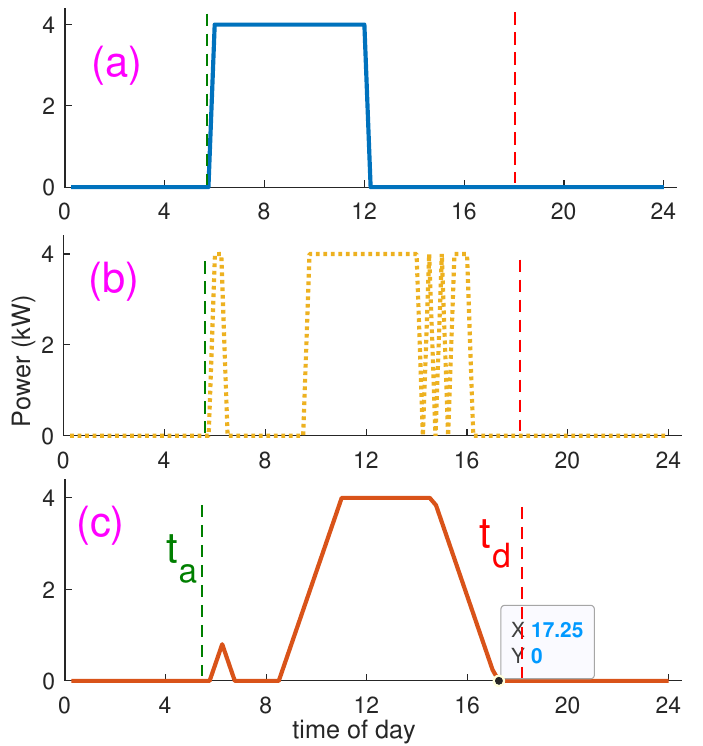}
    \vspace{-4pt}
    \caption{\small{Charging profiles comparison. (A) The nominal charging profile, (B) the charging profile based on smart charging without ramp rate constraint, (C) the charging profile for ramp rate constraint of is 10\% of the ramp (kW) limit.}}
    \label{fig:case2_res1}
\end{figure}

The flexibility is modelled via an electric vehicle. The electric vehicle is connected to the charger at 6 a.m. and leaves at 6 p.m. The connection time, thus, is 12 hours. However, the charging duration is only 6 hours and 15 minutes. The total charged energy is 25 kWh.
The rated charging power for this example is set to 4 kW.
Fig. \ref{fig:case2_res1}(a) shows the nominal charging profile.
The dashed green and red line shows the arrival and departure times.
Fig. \ref{fig:case2_res1}(b) shows the charging profile without any ramp rate constraint. Note that the charging modes are changed much more frequently and could lead to excessive switching for small cost savings.
Fig. \ref{fig:case2_res1}(c) shows the charging trajectory for the ramp rate constraint set to a mere 10\% of the rated charging power.

\begin{figure}[!htbp]
    \centering
    \includegraphics[width=13cm]{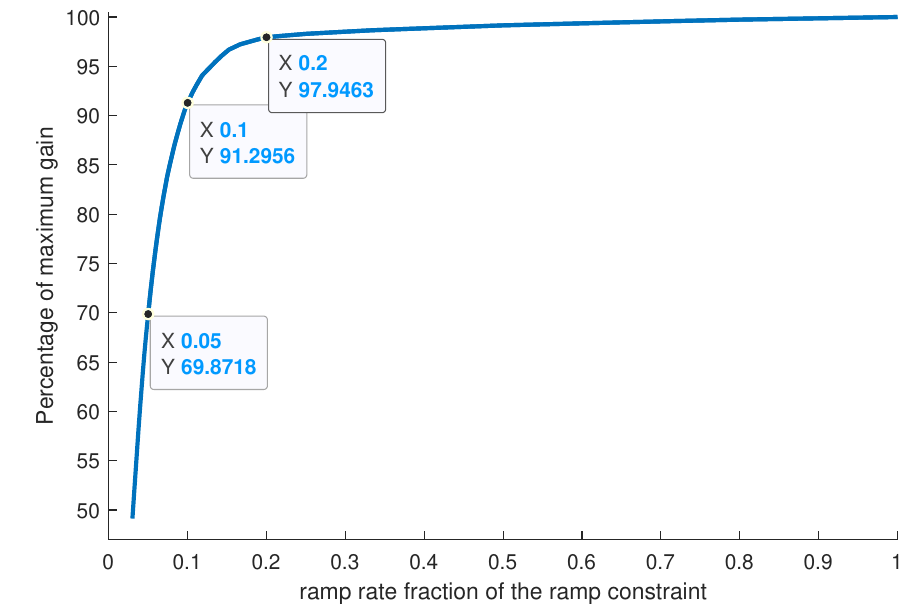}
    \vspace{-4pt}
    \caption{\small{Marginal cost savings (profit) for EV charging with varying ramp rate constraint levels.}}
    \label{fig:case2_res2}
\end{figure}

Similar to energy storage results, we observe that the marginal energy consumption cost savings are substantially high for low levels of ramp rate. 
Fig. \ref{fig:case2_res2} shows that for a ramp rate limit of 10\%, up to 91\% of cost savings can be achieved compared to the case where no ramp rate constraint is considered. This is very encouraging for flexibility sources with slow ramp rate characteristics.

\pagebreak

\section{Conclusion}
\label{section5}
Unlocking the full potential of energy storage and flexibility requires advanced modelling techniques that accurately represent these assets. Without such models, the true capabilities of these resources are often misjudged. Our research introduces an innovative energy storage and flexibility model that comprehensively accounts for ramp and capacity limits while precisely incorporating the ramp rate constraint. These models are linear and can be efficiently solved using standard linear programming solvers. To support wider application and benchmarking, we have created an online repository of these models.

Through detailed numerical case studies, we examine the impact of ramp rate constraints on the profit maximization goals for energy storage and flexibility. The results are particularly promising for assets with slower ramp rates. Our findings indicate that resources with a ramp rate limit set at 10\% of the maximum ramp limit can achieve between 65\% and 90\% of the maximum possible profit compared to scenarios without ramp rate constraints. This demonstrates the significant value and potential of accurately modelled energy storage and flexibility resources, even under restrictive ramp rate conditions.

 \pagebreak

\section*{Online repository}
The code developed for LP-based storage and flexibility models including the ramp-rate constraint described in this paper are publicly available at: 
\url{https://github.com/umar-hashmi/FlexibilityAndStorageModel}.
The online repository contains simulation files in MATLAB, Python and GAMS.

\section*{Acknowledgement}
This work is supported by the 
Flemish Government and Flanders Innovation \& Entrepreneurship (VLAIO) through the Flux50 projects InduFlexControl (HBC.2019.0113), and project
IMPROcap (HBC 2022.0733).

 \pagebreak

\bibliographystyle{IEEEtran}
\bibliography{reference.bib}

\end{document}